\documentclass[superscriptaddress,groupedaddress,nofootnoteinbib,11pt]{article}
\pdfoutput=1 
\usepackage{jheppub}

\usepackage{CJKutf8} 
\usepackage{mathtools,slashed,mathrsfs}
\usepackage[caption=false]{subfig}
\usepackage{dcolumn}
\usepackage{multirow}
\usepackage{tabularx}
\usepackage{booktabs}
\usepackage{bm}
\usepackage{amsmath}
\usepackage{comment}
\usepackage{setspace}
\usepackage[dvipsnames]{xcolor}
\usepackage[normalem]{ulem} 
\usepackage{enumerate}
\usepackage{siunitx}
\usepackage{float}
\usepackage{hyperref}

\definecolor{mypurple}{RGB}{164,64,214}

\newcommand\eea{\end{eqnarray}}
\newcommand\bea{\begin{eqnarray}}

\newcommand\bes{\begin{split}}
\newcommand\ees{\end{split}}
\newcommand{\mA}{m_A}
\newcommand{\A}{A_D}
\newcommand{\mphi}{m_{\phi}}

\begin{document}

\title{Adiabatic Conversion of ALPs into Dark Photon Dark Matter}

\date{\today}
\author[a]{Edward Broadberry,}
\author[b]{Saurav Das,}
\author[a]{Anson Hook,}
\author[c]{Gustavo Marques-Tavares}

\affiliation[a]{Maryland Center for Fundamental Physics, University of Maryland, College Park, MD 20742}
\affiliation[b]{Department of Physics and McDonnell Center for the Space Sciences,
Washington University, St. Louis, Missouri 63130, USA}
\affiliation[c]{Department of Physics and Astronomy, University of Utah, Salt Lake City, UT 84112, U.S.A.}

\emailAdd{EdBroad@umd.edu}
\emailAdd{s.das@wustl.edu}
\emailAdd{hook@umd.edu}
\emailAdd{g.marques@utah.edu}

\abstract{
 
We introduce a mechanism by which a misaligned ALP can be dynamically converted into a dark photon in the presence of a background magnetic field. An abundance of non-relativistic ALPs will convert to dark photons with momentum of order the inhomogeneities in the background field; therefore a highly homogeneous field will produce non-relativistic dark photons without relying on any redshifting of their momenta. Taking hidden sector magnetic fields produced by a first order phase transition, the mechanism can reproduce the relic abundance of dark matter for a wide range of dark photon masses down to $10^{-13}$ eV.
}

\maketitle

\section{Introduction}

Much is unknown about the properties of dark matter.  A very intriguing option is that dark matter is a light vector particle~\cite{Nelson:2011sf,Arias:2012az,Hui:2016ltb,Redondo:2008ec}, typically referred to as a dark photon.  
Dark photons are well motivated for both theoretical reasons, they often appear in string theory examples~\cite{Goodsell:2009xc,Cicoli:2011yh}, and phenomenological reasons, such as mediators of interactions to a dark sector~\cite{Pospelov:2008jk,Hall:2009bx,Essig:2011nj,Knapen:2017xzo}. 
There are a wide range of proposed experiments to search for dark photon dark matter, such as direct detection experiments, e.g. \cite{Expt1,Expt3,Expt4,Expt5,Arvanitaki:2017nhi,Expt6,Expt7,Fedderke:2021aqo,Jiang:2023jhl,Bloch:2023wfz,An:2013yua,An:2014twa,DAMIC:2016qck,Hochberg:2016ajh,Hochberg:2016sqx,Bloch:2016sjj,ADMX:2010ubl}, as well as many indirect astrophysical searches, see e.g. \cite{Dubovsky:2015cca,Expt8,Expt9,Witte:2020rvb,An:2013yfc,Redondo:2013lna,Hardy:2016kme,Irsic:2017yje,Baryakhtar:2017ngi,Cardoso:2017kgn,Cardoso:2018tly,Fukuda:2018omk}. For a review of experiments related to dark photon dark matter consult \cite{Essig:2013lka}. Despite their rich phenomenology, and the large experimental effort searching for light vector dark matter, there are only a limited number of cosmological production mechanisms that can achieve the correct dark matter abundance. 

Interestingly, most of these new experiments are dominantly sensitive to dark photons with very light masses so that they behave as coherent fields.  This field-like regime occurs when the mass of dark matter is between eV $\gtrsim m \gtrsim 10^{-19}$ eV, with eV being dictated by when each dark matter particle's de Broglie wavelength starts to overlap, and the $10^{-19}$ eV bound comes from galactic dynamics~\cite{Dalal:2022rmp}. It is very difficult to produce dark photon dark matter at the smallest of masses. The reason for this is that the dark photon field must point in a direction. Producing dark photons completely at rest would involve breaking rotation invariance. Since all models producing dark matter respect rotation invariance, the directions of the produced dark photon field must average to zero over large volumes. Therefore, they cannot be fully homogeneous and the produced dark photons must have a velocity. Reconciling this non-zero momentum with dark matter's non-relativistic nature is why producing very light dark photon dark matter is difficult.

This non-zero momentum manifests itself in every dark photon dark matter production mechanism. When produced from inflationary fluctuations \cite{Graham:2015rva,VectorDMProduction} or cosmic strings~\cite{CosmicStrings} dark photons are created with momentum of order Hubble. Parametric resonance~\cite{ParametricResonance} produces dark photons with momentum of order the mass of the decaying scalar and thus requires the two masses tuned to be close to get non-relativistic dark matter. There are many other production mechanisms, all of which have to deal with this tension in some way or another. A popular category of production mechanisms, and the one we will pursue, is to use the misalignment mechanism to populate some dark scalar, and find a way to convert this energy into a dark photon \cite{RelicAbundance,ParametricResonance,AxionOscillations}. 

We take inspiration from recent work on the mixing of a dark photon with an ALP \cite{Hook:2019hdk}. It was found that in the presence of a background magnetic field, mixing effects between a dark photon and an ALP can decrease Hubble friction making the energy density of an ALP decay as $a^{-1}$ instead of the usual $a^{-3}$.
We will use this same phenomenon to convert an ALP into dark photon dark matter.
We will consider an ALP coupled as~\cite{Kaneta:2016wvf,Kaneta:2017wfh,Pospelov:2018kdh,Choi:2018mvk,Kalashev:2018bra,Biswas:2019lcp,Choi:2019jwx,Hook:2019hdk,deNiverville:2020qoo,Arias:2020tzl,Hook:2021ous,Hook:2023smg}
\bea
\mathcal{L}\supset \frac{\phi}{f}F\tilde{F}_d,
\eea
which describes an interaction between an axion like particle (ALP), a dark photon and another photon (that may or may not be the SM photon). In the presence of a magnetic field, this interaction will adiabatically convert cold ALPs into cold dark photons.

Our particular cosmology proceeds as follows. A first order phase transition will create a turbulent charged plasma, which by equipartition generates long range coherent magnetic fields. This large magnetic field causes large mixing between the dark photon and the ALP. At some point later, the misaligned ALP begins to roll, in the process converting back and forth with the dark photon.  Eventually, the magnetic field is too small to keep up the conversion between the ALP and the dark photon and all of the energy is transferred into the lighter dark photon.  In this manner, the cold ALP is converted into a cold dark photon.  Using our mechanism, dark photons with masses as small as $10^{-13}$ eV can be produced in the magnetic field from a hidden sector $U(1)$. 

In our approach we will be taking the phase transition, turbulent charged plasma, and long range coherent magnetic fields to all be part of a separate dark sector because it is unclear if conversion in the SM electromagnetic field is viable. This is due to the presence of charged particles with masses below the temperature of the phase transition wreaking havoc with the coherence length of the B-field. The modeling of such a scenario would require dedicated numerical simulations in an expanding universe. We discuss potential limits for this scenario in appendix A.


In section \ref{sec: review}, we review previous results.  In section \ref{sec:homogeneous}, we discuss conversion of an ALP into a dark photon in the presence of a homogeneous magnetic field.  Section \ref{sec: inhomo} treats the conversion in an inhomogeneous magnetic field.  Finally, we conclude in section \ref{sec: concl}.

\section{Review of Previous Work} \label{sec: review}

In previous work \cite{Hook:2019hdk}, it was shown that a large mixing between an axion like particle (ALP) and a dark photon induced by a background magnetic field can parametrically enhance the late-time abundance of the ALP. The Lagrangian under consideration is 
\begin{equation}
    \label{eq:Lagrangian}
    \hat{\mathcal{L}} = \hat{\mathcal{L}}_{\rm kin}(\phi,A,\A) - \frac{1}{2}m_\phi^2\phi^2 + \frac{1}{2}m_A^2 A_{d,\mu}A_d^\mu + \frac{1}{\sqrt{-g}}\frac{\phi}{2f}F_d\tilde{F},
\end{equation}
where the hat indicates the factoring out of $\sqrt{-g}$. In the presence of a homogeneous background magnetic field, $\vec B(t)$, the equations of motion in an FLRW universe with scale factor $a(t)$ are
\begin{align}
        \ddot{\phi} + 3 H \dot{\phi} + \mphi^2 \phi &= \frac{b}{a}\dot{A}_D,\nonumber \\
        \ddot{A}_D+H\dot{A}_D+\mA^2 \A &= - a b \dot{\phi},\label{eq:EOM1}
\end{align}
where we have defined the mixing, $b(t) = B(t)/f$, $\A(t)$ is the component of the dark photon field parallel to $\vec B$ in $\partial_\mu \A^\mu = 0$ gauge, and $H = \dot{a}/a$ is the Hubble expansion rate. The ALP is produced via the misalignment mechanism, while the initial vector abundance after inflation is taken to be negligible, so that the initial conditions for both fields are
\begin{align}
\label{eq:cosmoinitial1}
    \phi(t_{\rm PT})  = \phi_{\rm PT}, & \ \ \ \dot{\phi}(t_{\rm PT}) = 0,\\
\label{eq:cosmoinitial2}
     \A(t_{\rm PT})  = & \ \dot{A}_D(t_{\rm PT}) = 0,
\end{align}
where we have denoted the initial time as $t_{\rm PT}$ for consistency with a later section where the initial time is identified with a first order phase transition.

It is useful to analyze the system in terms of the instantaneous normal modes in the friction-less limit (i.e. ignoring Hubble friction). To gain more intuition, we also first take $b$ to be time independent. Focusing first on the regime where $m_A \gg m_\phi$, we find that the two oscillation frequencies of the system are:
\begin{align}
    \omega_s^2 &\approx \frac{m_\phi^2 m_A^2}{b^2+m_\phi^2+m_A^2},\label{eq:slowfreq}\\
    \omega_f^2 &\approx b^2 + m_\phi^2 + m_A^2.\label{eq:fastfreq}
\end{align}
In the large mixing regime, i.e. $b \gg m_A, \, m_\phi$, the first frequency is much smaller than either of the masses, while the second one is much larger. We will denote the normal mode associated with the smaller frequency the slow mode, and the one associated with the larger frequency the fast mode.

The slow mode satisfies
\begin{equation}
\label{eq:SlowVec}
    \A \approx i \frac{m_\phi}{m_A}\frac{b}{\sqrt{b^2+m_A^2}} \, \phi \, ,
\end{equation}
whereas the fast mode,
\begin{equation}
    \phi \approx i \frac{b}{\sqrt{b^2 + m_A^2}} \, \A \, ,
\end{equation}
where in both expressions we have used $m_A \gg m_\phi$, but did not assume $b > m_A$.
Note that when $b \gg m_A$, the vector field's amplitude is suppressed by $m_\phi/m_A$ compared to the scalar for the slow mode while for the fast mode both amplitudes are approximately the same. However, as the mixing goes to 0 the slow mode becomes the lighter field, $\phi$, and the fast mode becomes the heavier field, $\A$. Given the initial conditions in Eqs.~\eqref{eq:cosmoinitial1} and~\eqref{eq:cosmoinitial2}, if the mixing is large, $b_{\rm PT}\gg m_A$, the initial amplitude for the slow mode goes as $\phi_{\rm PT} \, b_{\rm PT}^2/(b_{\rm PT}^2+m_\phi^2)$, whereas the amplitude for the fast mode goes as $\phi_{\rm PT} \, m_\phi^2/(b_{\rm PT}^2+m_\phi^2)$. Thus, because the initial condition has zero kinetic energy, the fast mode is effectively not excited in the $b \gg m_\phi$ limit.




Much of the intuition from the previous discussion carries over by considering the instantaneous normal modes of Eq.~\eqref{eq:EOM1} after allowing for time depedence by including the expansion $H$ and $\dot{b}\neq0$. The equations of motion \eqref{eq:EOM1} resemble harmonic oscillators with time dependent friction. In analogy with the misalignment mechanism we define the crossing time, $t_X$, as the time at which the slow mode becomes underdamped 
\begin{equation}
    \label{eq:CrossingTime}
    H_X = \omega_{s}(t_X)
\end{equation}
where a subscript, $X$, indicates a quantity evaluated at crossing time. Notice that if the mixing is large, $b_X > m_A$, the crossing time is later than in the usual misalignment mechanism because $\omega_s < m_\phi$. The fields are therefore frozen for longer, leading to a greater late time abundance. 

For a large mass hierarchy, the kinetic and friction terms for the dark photon are subdominant and we can approximate $m_A^2 \A \approx - a(t)b(t)\dot{\phi}$. Plugging this into equation \eqref{eq:EOM1} leads to
\begin{equation}
    \label{eq:EOM2}
    \ddot{\phi} + \left((3-2\lambda)H + \lambda \frac{\dot{b}}{b}\right)\dot{\phi} = -\omega_s^2(t)\phi,
\end{equation}
where $\lambda = b^2/(b^2+m_\phi^2+m_A^2)$. It was found that if the mixing changes adiabatically, the fields will approximately remain in the instantaneous slow mode. We can approximate the field as completely frozen for $t< t_X$.  Meanwhile for $t > t_X$, we use the ansatz
\begin{equation}
    \phi(t) \approx \varphi(t)\exp\left(i\int_{t_X}^t\omega_s(t')dt'\right)\label{eq:Ansatz1}.
\end{equation}
The time evolution of the amplitude $\varphi$ can be found using the WKB approximation and equation \eqref{eq:EOM2}. The solution exhibits a phenomena coined `gliding' \cite{Hook:2019hdk} and is given by
\begin{equation}
    \label{eq:Gliding1}
    \varphi(t) \approx \varphi_X \exp\left[-\frac{1}{2}\int_{t_X}^tdt'H\left(1+\frac{2m_A^2}{m_A^2 + b(t)^2}\right)\right].
\end{equation}
As a function of time the mixing will start large, $b_{\rm PT} \gg m_A$, but at late times $b(t)\rightarrow 0$. If the mixing is still large after $t_X$, then equation \eqref{eq:Gliding1} indicates $\phi \sim a^{-1/2}$, and thus the energy density decreases as $a^{-1}$. Once $b \ll m_A$, the ALP will start diluting like matter $\phi \sim a^{-3/2}$ (and, thus, $\rho \propto a^{-3}$), as expected since in this limit the mixing becomes negligible and the slow mode becomes effectively just the scalar field.

In summary, the previous work found that kinetic mixing with a heavy vector induced by a magnetic field enhances the abundance of ALPs by two effects. Firstly the ALP is frozen for longer, and thus its energy density begins diluting later. In addition to this there is a period in which the scalar `glides'; its energy density, $\rho_\phi$, dilutes more slowly than matter, $\rho_\phi \propto a^{-1}$. In this paper we invert the hierarchy $m_A \ll m_\phi$, and notice that at late times the slow mode is the dark photon. We will see that misaligning an ALP will still excite the slow mode, and the energy density will adiabatically convert from ALP to dark photon, leaving a late-time abundance of dark photons.

\section{Producing Dark Photons in a Homogeneous Background} \label{sec:homogeneous}

We first consider the simplified example of how the scalar misalignment mechanism (Eqs.~\eqref{eq:cosmoinitial1} and \eqref{eq:cosmoinitial2}) produces dark photons in the presence of a homogeneous magnetic field $b=b(t)$ when $m_\phi \gg m_A$.  The initial conditions still excite the slow mode, so long as the magnetic field evolves adiabatically we can assume almost all the final energy will remain in the slow mode. 

In the large mass limit, $\ddot \phi$ and $H \dot{\phi}$ are small compared to the mass term so that $m_\phi^2 \phi \approx b\dot{A}_D/a$. Plugging into the equations of motion gives an analogous result to equation \eqref{eq:EOM2} 
\begin{equation}
    \label{eq:GlidingEOM2}
    \ddot{A}_D+\left[(1-2\lambda)H+\lambda\frac{\dot{b}}{b}\right]\dot{A}_D = -\omega_s^2(t) \A.
\end{equation}
For a B-field that decays as a power law, $b \propto a^{-n}$, we can find an analytic solution in the WKB approximation,
\begin{eqnarray}
\nonumber
        \A(t) && = \mathcal{A}(t)\exp\left(i\int \omega_s(t')dt'\right)\\
        && \implies \mathcal{A}(t) = \mathcal{A}_X \sqrt{\frac{a}{a_X}}\left(\frac{b^2+m_\phi^2\left(\frac{a_X}{a}\right)^{2n}}{b^2 + m_\phi^2}\right)^{1/2n},
        \label{eq:HomGliding}
\end{eqnarray}
where $\mathcal{A}_X$ is the amplitude, and $a_X$ is the scale factor, evaluated at crossing time (i.e., when $\omega_s(t_X) = H(t_X)$). This solution is a very good approximation for times $t>t_X$. We study the case $b_X > m_\phi$, leaving the other ordering for the appendix. The energy density in the vector is given by 
\begin{equation}
\label{eq:homrho}
    \rho_{\A} \approx \frac{m_A^2}{2a^2}\mathcal{A}^2(t).
\end{equation}
The amplitude \eqref{eq:HomGliding} exhibits the same qualitative behaviour as previous work. For large mixing, $b \gg m_\phi$, the amplitude, $\mathcal{A} \propto a^{1/2}$, and the energy density falls off as $\rho_{\A} \propto 1/a$, so we see that the gliding mechanism persists when the vector is the lighter field. Finally, as the mixing turns off, $b\ll m_\phi$, the amplitude falls off as $\mathcal{A} \propto a^{-1/2}$, and the vector dilutes like matter $\rho_{\A} \propto a^{-3}$.

\subsection{Estimating the amplitude}

\begin{figure}[t]
  \centering
   \subfloat{\includegraphics[width=0.48\textwidth]{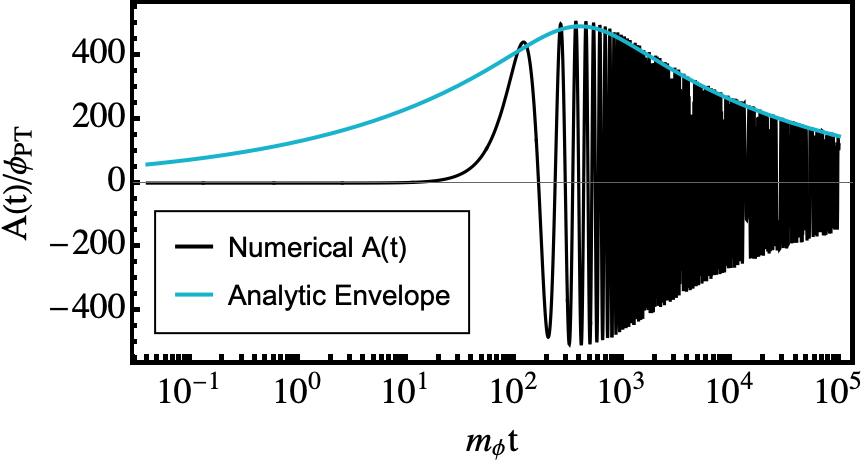}
  }
  \hfill
  \subfloat{\includegraphics[width=0.46\textwidth]{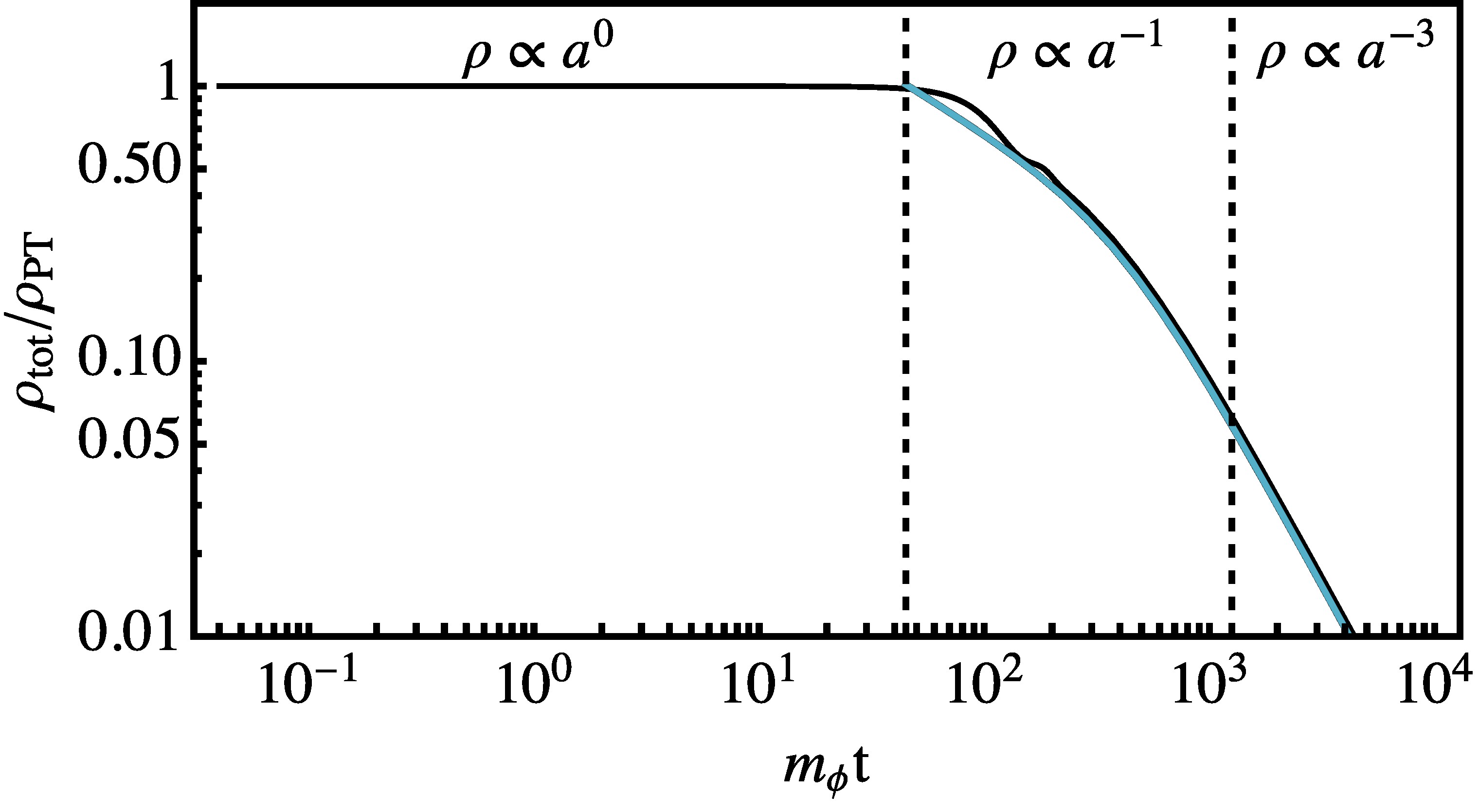}}
  \hfill
  \caption{Left: In black we plot the numerical solution to equation \eqref{eq:EOM1} with parameters chosen to satisfy the gliding criteria, $b_X>m_\phi$. In teal we plot equation \eqref{eq:HomGliding} with $\mathcal{A}_X$ given by equation \eqref{eq:Ax}, and the crossing time defined by \eqref{eq:CrossingTime}. We see that the analytical estimate closely approximates the amplitude of the vector field. Right: In black we plot the total energy, $\rho_\phi + \rho_A$. In teal we plot equation \eqref{eq:homrho}, with $\mathcal{A}(t)$ given by the line on the left graph. The agreement shows that all the energy density of the system is in the vector.}
    \label{fig:HomGliding}
\end{figure}

The amplitude of the vector can be determined within $\mathcal{O}(1)$ factors by assuming that all the energy is in the slow mode. We note that the fields are frozen before $t_X$, so we can assume $\phi_X \sim \phi_{\rm PT}$, and then use the equivalent of equation \eqref{eq:SlowVec} with the inverted mass hierarchy to find
\begin{equation}
\label{eq:Ax}
    \mathcal{A}_X \approx \frac{m_\phi}{m_A} \frac{\sqrt{b_X^2+m_\phi^2}}{b_X}a_X \phi_{\rm PT}.
\end{equation}
In figure, \ref{fig:HomGliding} we plot both the numerical solution in black and the analytic solution for the envelope from \eqref{eq:HomGliding}, and \eqref{eq:Ax} in blue.  We see that the analytic estimate matches the numerical solution to the equations of motion \eqref{eq:EOM1} when parameters are chosen such that the mixing is still large at the crossing time. 

\begin{figure}[ht]
\centering
	\includegraphics[width=0.7\linewidth]{"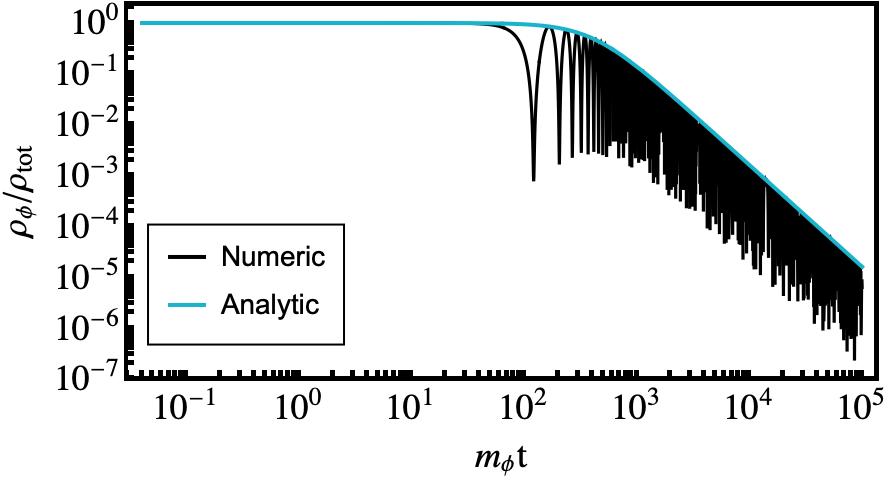"}
	\caption{In black we plot the fraction of the energy density in the scalar $\rho_\phi/\rho_{\rm tot}$ for the numeric solution to equation \eqref{eq:EOM1}. In teal we plot equation \eqref{eq:conversion} and find good agreement. For the entire gliding parameter space, that the approximation that all the energy density goes into dark photons is well justified, since equation \eqref{eq:LZ} is always satisfied.}
	\label{fig:HomConversion}
\end{figure}

At late times the B-field dilutes away and the slow mode becomes the vector field $\A$. Effectively, the conversion from ALP to dark vector occurs at $t_M$, defined by
\begin{equation}
    b(t_M) = m_\phi.
    \label{eq:def:TM}
\end{equation}
As long as $m_\phi/H_M \gtrsim 1$ the conversion efficiency is exponentially sensitive to the ratio $m_\phi/H_M$, where the subscript $M$ indicates quantities evaluated at $t_M$, with
\begin{equation}
    \frac{\rho_\phi}{\rho_{\rm tot}} \propto \exp\biggl(- c \frac{m_\phi}{H_M}\biggr),
\end{equation}
for some constant $c$. The exponential sensitivity comes from an analogy with the Landau-Zener formula, which is expanded on in appendix \ref{app:homogeneous constraints} \cite{Landau,Zener}. The conversion from ALP to dark photon is therefore efficient provided that
\begin{equation}
    \label{eq:LZ}
    \frac{m_\phi}{H_M} \gg 1.
\end{equation}
The condition in \eqref{eq:LZ} is always satisfied in the gliding regime, where $b_X > m_\phi$. Using the analytic solution \eqref{eq:HomGliding} we can estimate that at late times $\rho_\phi/\rho_A \sim b(t)^2/m_{\phi}^2$
\begin{equation}
    \frac{\rho_\phi}{\rho_{\rm tot}} \approx \frac{b(t)^2}{b(t)^2 + m_\phi^2}.
    \label{eq:conversion}
\end{equation}
The only reason why there is any energy at all in $\phi$ is that the slow mode oscillations still have some non-trivial overlap with the heavy scalar $\phi$.
In figure \ref{fig:HomConversion} we test equation \eqref{eq:conversion}.

The authors are not aware of an effective mechanism to create a sufficiently homogeneous (at the relevant scales) B-field of significant strength, so in the next section we begin studying the effect of inhomogeneities. However, in appendix \ref{app:homogeneous constraints} we show that if a mechanism for producing a very homogeneous magnetic field existed it would be possible to produce dark photon dark matter down to the lowest viable masses of $m_A \sim 10^{-19}$ eV.

\section{Producing Dark photons in a Inhomogeneous Magnetic Field} \label{sec: inhomo}

\subsection{How Magnetic Fields are generated and how they decay}
\label{sec:BFields}

Given the central role magnetic fields play in this production mechanism, and the fact that their inhomogeneities will lead to significant changes to the final dark photon abundance, we will first discuss an efficient production mechanism for large cosmological magnetic fields, and the subsequent evolution of the generated field.

While first order phase transitions are interesting for many reasons, one of their common features is the production of magnetic fields~\cite{Baym:1995fk,Grasso:1997nx,ENQVIST_1998}, which can lead to many interesting consequences~(see, e.g.~\cite{DeSimone:2011ek,Ellis:2019tjf}).  The random production of bubbles and their subsequent collisions generates turbulence in the surrounding plasma. This turbulent behavior will amplify any existing seed magnetic fields into large scale magnetic fields~\cite{stevens2010theory}.  An order of magnitude estimate of the size of these magnetic fields can be obtained by equipartition, $\rho_B \sim v_f^2 \rho_{\gamma}$, where $v_f$ is the typical turbulent velocity of the fluid and $\rho_\gamma$ is the energy density in the photon fluid.  A more detailed analysis reveals that the coefficients are typically order one and that the wavelength of the produced magnetic field is of order the inverse size of the bubbles~\cite{Baym:1995fk,Sigl_1997}.  As a result of this, the magnetic field generated by a first order phase transition can be easily as large as
\bea
B^2 \sim 0.1 T^4.
\eea

After production, the produced magnetic fields slowly decay.  If at any point the charged plasma vanishes, e.g. if the temperature goes below the mass of the lightest charged particle, then the magnetic field just decays away as $a^{-2}$ while the characteristic momentum decays away as $a^{-1}$.  If the turbulent charged plasma is still present, then the magnetic fields will discharge faster than just expansion of the universe.  How the magnetic fields decay due to turbulence has been mainly studied numerically~\cite{PhysRevLett.83.2195,M_ller_2000,Boyarsky:2011uy,Durrer:2013pga}.
These numerical results find the scalings in equations \eqref{eq:lamba}, \eqref{eq:b}, \eqref{eq:blambda}.
However there are a few rules of thumb that roughly reproduce what is seen in these simulations that we will discuss below.  For simplicity, in the following paragraphs we will discuss the discharge of magnetic fields in the absence of Hubble expansion.

The decay of magnetic fields depends heavily on whether the produced magnetic fields are helical or non-helical. 
Non-helical magnetic fields are the easiest to understand.  Due to equipartition of energy
\bea
\rho_\gamma v_f^2 \sim B^2 .
\eea
If we take $\lambda$ to be the typical length (inverse momentum) scale of the problem, then because CP and P are not broken, correlation lengths cannot do anything but a random walk, so that
\bea \label{eq:lamba}
\lambda \propto \sqrt{t}.
\eea
Combining this with $\lambda = v_f(t) t$, we find that the magnetic fields decay as
\bea \label{eq:b}
B \propto \sqrt{\frac{\rho_\gamma(t)}{t}}
\eea

On the other hand, helical Magnetic fields are subject to conservation of magnetic helicity
\bea
\partial_t ( A \cdot B ) = -2 E \cdot B + \text{total derivatives} = - 2 \eta J \cdot B,
\eea
where we have used Ohm's Law $ E = -v \times B + \eta J$, where $\eta$ is the resistivity.  Plasmas are good conductors so $\eta \approx 0$ and magnetic helicity is conserved.
If we take $\lambda$ to be the inverse of the typical momentum scale of the problem, then conservation of magnetic helicity ($A \cdot B \propto \lambda B^2$) means that
\bea \label{Eq:lambda}
B \propto \frac{1}{\sqrt{\lambda}}.
\eea
As before, equipartition of energy between plasma and B fields gives $\rho_\gamma v_f^2 \sim B^2$ while $\lambda \sim v_f t$.  Putting it all together gives
\bea \label{eq:blambda}
B \propto \frac{\sqrt{\rho_\gamma}}{t^{1/3}} \qquad \lambda \propto t^{2/3}.
\eea
Because we did not specify the dependence of Eq.~\eqref{Eq:lambda} on $\rho_\gamma$, the dependence of Eq.~\eqref{eq:blambda} on $\rho_\gamma$ requires additional comment. 
Fixing $B^2 \sim \rho_\gamma$ at $t_{PT}$ implies the proportionality $B \propto \sqrt{\rho_\gamma}$.

\subsection{Dynamics in an inhomogenous magnetic field}


In section 3 it was shown that an initial abundance of an ALP can be efficiently converted into dark photons provided the ALP is the heavier particle, $m_\phi \gg m_A$, and the mixing is large at the crossing time \eqref{eq:CrossingTime}, $b_X > m_\phi$. However, in a phase transition the generated magnetic fields are not correlated at scales larger than the horizon size at that time (generically even on smaller scales), and thus are not homogeneous over large scales, but have a characteristic correlation scale $k$ \cite{Baym:1995fk,Grasso:1997nx,ENQVIST_1998}. In order to understand the effects of inhomogeneities in the conversion mechanism,  we consider the following simplified field profile
\begin{equation}
    \label{eq:InhomogeneousField}
    \vec{b}(x,t) = b_{\rm PT}\left(\frac{a_{\rm PT}}{a}\right)^2\cos (kx) \hat{z},
\end{equation}
for comoving coordinate $x$. A field with the profile in \eqref{eq:InhomogeneousField} does not really correspond to a realistic cosmological magnetic field characterized by a correlation scale $k$ for several reasons. The first is that the direction is the same everywhere in the universe, a large breaking of rotational invariance.  Any causal production mechanism would preserve rotational invariance on super-horizon scales, and it is expected that for points separated by scales larger than $1/k$ the direction and amplitude of the field would be largely uncorrelated. The second is that the correlation scale, $k$, and amplitude of the B field only evolves due to the expansion of the universe.  As discussed in the previous subsection, the presence of charged particles with masses below the phase transition temperature wreaks havoc with the coherence length and amplitude of the B-field. To avoid dealing with the extra complications arising from a changing coherence scale, we instead consider the B-field to originate from a dark $U(1)$\footnote{It would be interesting to know if the conversion mechanism would survive a realistic magnetic field profile from our sector, but the analysis would require a more sophisticated numerical modeling beyond the scope of this work.} with all charged particles being sufficiently heavy, $T_X \leq M \leq T_{\rm PT}$. where $T_{\rm PT}$ is the temperature of the phase transition and $M$ the charged particle mass. This ensures the charged particles are around to generate the magnetic field, but there are no plasma dynamics to affect the conversion. In this case inhomogeneities of size $k$ will dilute as \eqref{eq:InhomogeneousField}. The equations of motion are as follows
\begin{align}
    \ddot{\phi}+3H\dot{\phi}+m_\phi^2 \phi - \frac{1}{a^2}\phi'' &= \frac{b}{a}\dot{A}_D,\nonumber\\
    \ddot{A}_D+H\dot{A}_D+m_A^2 \A - \frac{1}{a^2}\A'' &= -ba\dot{\phi},
    \label{eq:EOM3}
\end{align}
where a prime indicates differentiation with respect to $x$. To solve these equations numerically we place the fields in a box of size $2\pi/k$ and impose periodic boundary conditions.

\subsection{Effective Mass approximation}
\label{sec:Gliding}
\begin{figure}[t]
  \centering
   \subfloat{\includegraphics[width=0.48\textwidth]{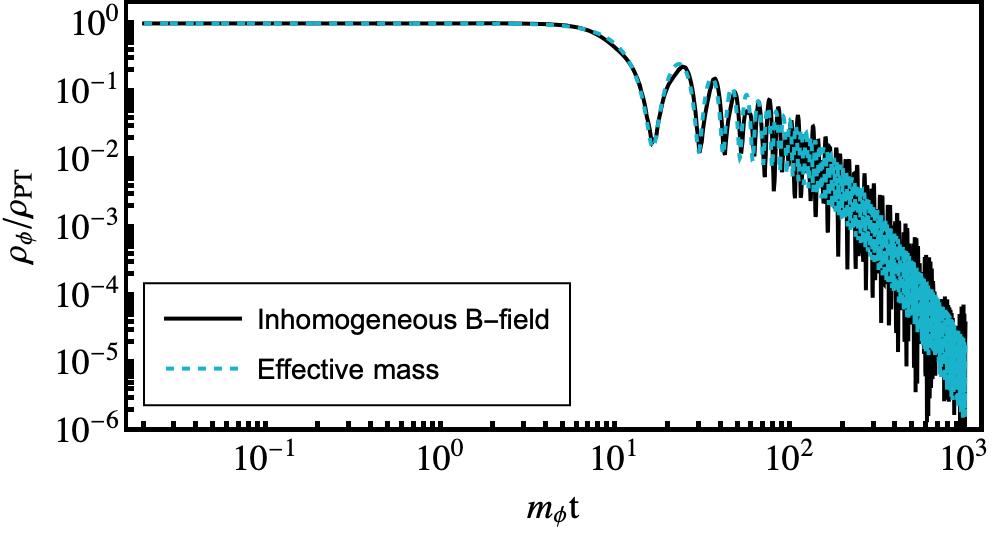}
  }
  \hfill
  \subfloat{\includegraphics[width=0.48\textwidth]{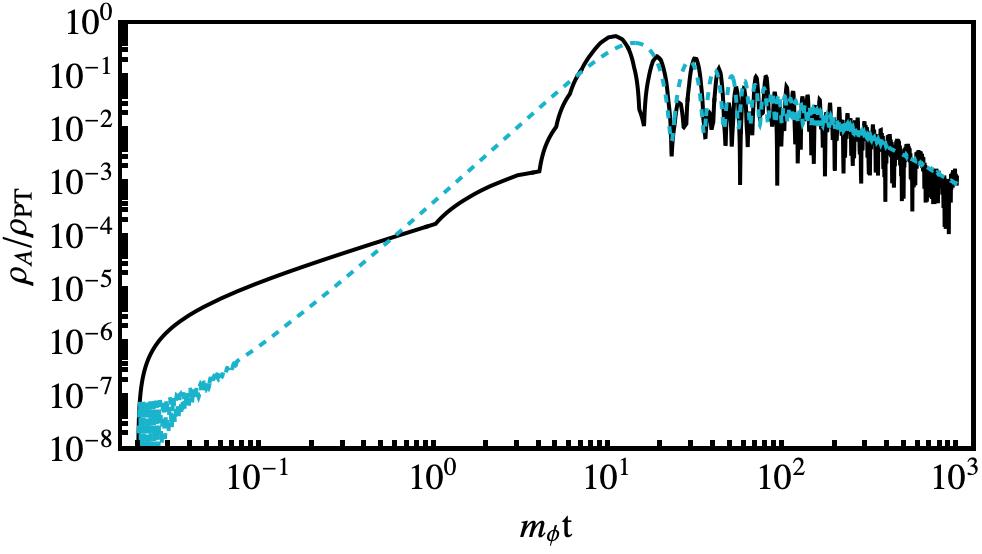}}
  \hfill
  \caption{Above is a plot of the energy densities for the scalar and vector as a function of time. The black line is the numerical solution to \eqref{eq:EOM3} in a box of size $2\pi/k$ with periodic BCs. The blue line is the numerical solution to \eqref{eq:EOM1} with the time-dependant mass given in equation \eqref{eq:meff}. The parameters were $k=10m_\phi$, $m_\phi = 10m_A$, $b_{\rm PT} = 10^4m_\phi$, and $H_{\rm PT} = 25m_\phi$.}
  \label{fig:EffectiveMassDemonstration}
\end{figure}
In previous work \cite{Hook:2019hdk} it was shown that for $k \gg m_\phi$, we can effectively think of this as a two state system consisting of the zero mode of $\phi$, and the $k^{\rm th}$ Fourier mode of $\A$. In equation \eqref{eq:EOM3}, the spatial derivative acting on $\A$ will turn into a $k^2$, whereas it will give zero acting on $\phi$. The same analysis holds when the mass hierarchy is inverted and the spatial derivative will look like an effective  time-dependent mass for the dark photon. This suggests we model the inhomogeneity as a homogeneous b-field, with a time dependent mass
\begin{align}
    \nonumber b_{\rm eff}(t) &= \frac{b_{\rm PT}}{\sqrt{2}}\left(\frac{a_{\rm PT}}{a(t)}\right)^2,\\
    m_{\rm eff}(t) &= \sqrt{m_A^2 + k^2 \left(\frac{a_{\rm PT}}{a(t)}\right)^2}.
    \label{eq:meff}
\end{align}
There is a relative factor of $\sqrt{2}$ due to averaging over a cosine. We checked the effective mass approximation for the range of parameters of interest; Figure \ref{fig:EffectiveMassDemonstration} shows an example of the agreement.

\begin{figure}[t]
  \centering
   \subfloat{\includegraphics[width=0.48\textwidth]{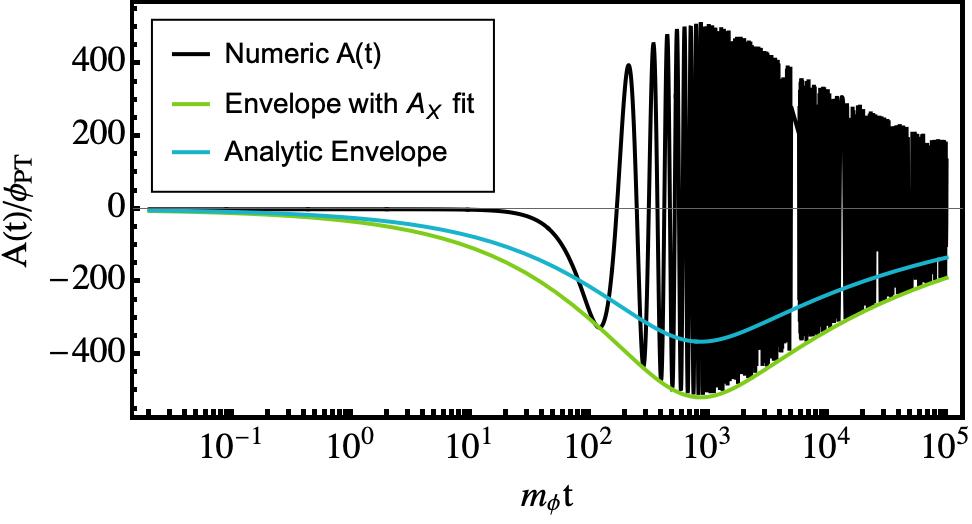}}
  \hfill
  \subfloat{\includegraphics[width=0.47\textwidth]{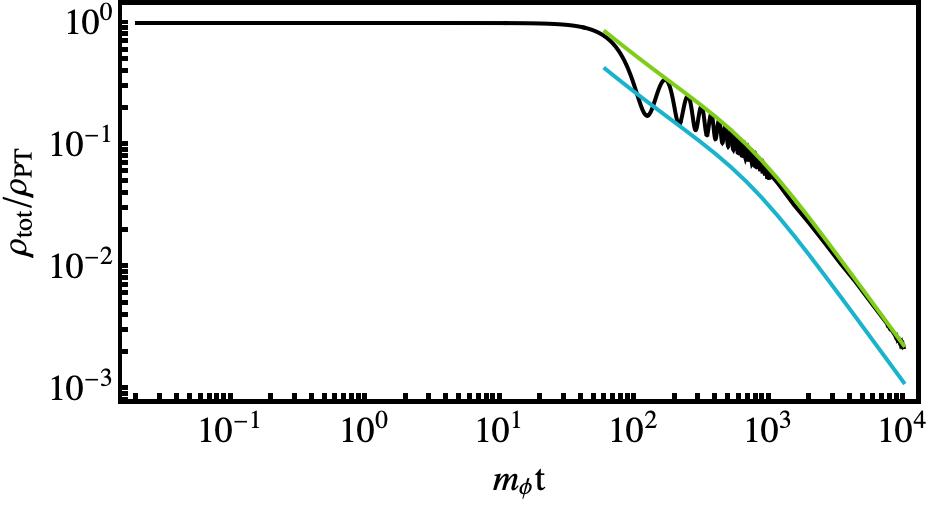}}
  \hfill
  \caption{A plot of vector field, and total energy density over time in the gliding regime. The numeric solution is shown in black. The green envelope is the analytic result in equation \eqref{eq:GlidingFormulaInh}, but with $\mathcal{A}_X$ fit to the numeric result, and the blue envelope is a fully analytic result with $\mathcal{A}_X$ given by equation \eqref{eq:Ax}.}
    \label{fig:InHomGliding}
\end{figure}

In the homogeneous case the gliding regime applied for $b_X > m_\phi$. We find that the gliding regime still applies provided the vector is the lighter field $m_{\rm eff}<m_\phi$ when the mixing becomes subdominant, $b=m_\phi$.  However, the presence of a time-dependent mass modifies the formula \eqref{eq:HomGliding} to \cite{Hook:2019hdk}
\begin{equation}
\label{eq:GlidingFormulaInh}
    \mathcal{A}(t) = \mathcal{A}_X \sqrt{\frac{m_{\rm eff}(t_X)}{m_{\rm eff}(t)}\frac{a(t)}{a_X}}\left(\frac{b^2+m_\phi^2\left(\frac{a_X}{a}\right)^{2n}}{b^2 + m_\phi^2}\right)^{1/2n}.
\end{equation}
We can try to estimate the amplitude $\mathcal{A}_X$ by equation \eqref{eq:Ax}, replacing $m_A \rightarrow m_{\rm eff}(t_X)$. The fit is not as good, but still gives the correct scaling. The scaling can be easily understood, since in this picture the homogeneous (0 momentum) axion converts to dark photons with comoving momentum $k$, and while those are relativistic, their energy density redshifts as radiation, but once $k a_{\rm PT}/a < m_A$, the dark photon becomes non-relativistic and its energy density starts redshifting like matter.
To estimate the final dark photon abundance, we can work at 3 levels: the fully numeric approach, the fully analytic approach with $\mathcal{A}_X$ given by equation \eqref{eq:Ax}, or a semi-analytic approach where we use equation \eqref{eq:GlidingFormulaInh}, but fit $\mathcal{A}_X$ to the numerics. In figure \ref{fig:InHomGliding} we show the matching between the numerics and various analytic approaches for the amplitude of $A$. 
Approximating all the energy density as being in the vector,
\begin{equation}
    \rho_{\rm tot} \approx \frac{1}{2a^2(t)}m_{\rm eff}^2(t)\mathcal{A}^2(t).
\end{equation}
The right hand side of figure \ref{fig:InHomGliding} also shows agreement between the numeric and analytic approaches for the energy density, where the analytic lines begin at the crossing time. The addition of $m_{\rm eff}(t)$ has no effect on equation \eqref{eq:conversion}, since the amplitude, $\mathcal{A}(t)$ cancels in $\rho_\phi/\rho_A$. We checked numerically that equation \eqref{eq:conversion} accurately describes energy transfer from the ALP to the dark photon, and the graph looks identical to figure \ref{fig:HomConversion}.

\subsection{Constraints}

\begin{figure}[t]
\centering
	\includegraphics[width=0.7\linewidth]{"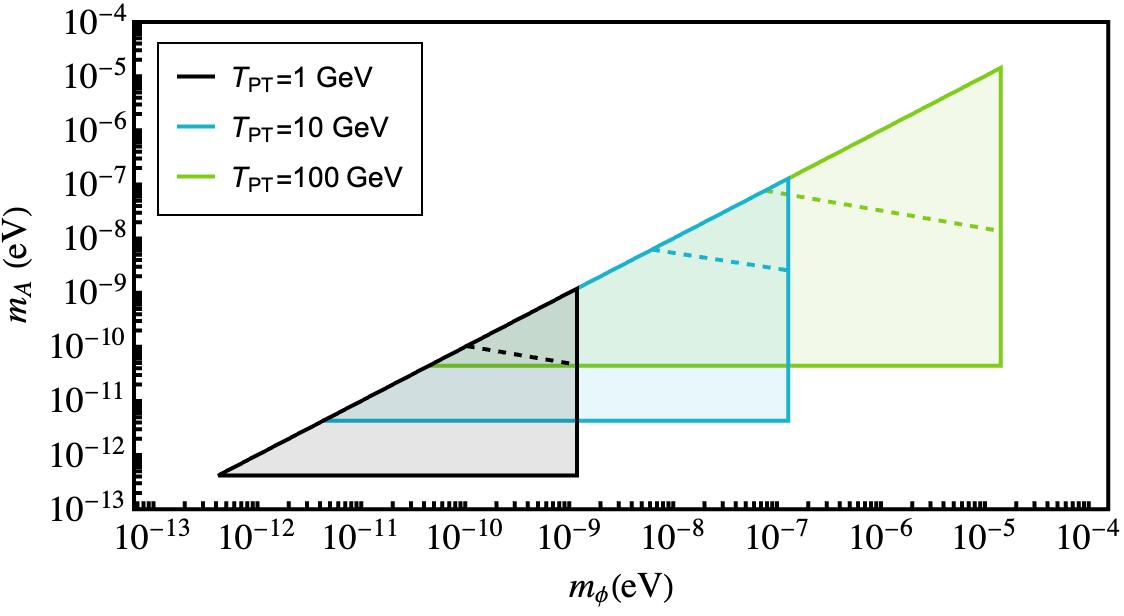"}
	\caption{The allowed region of parameter space for phase transitions occurring at $T_{\rm PT} = 1,10,100$ GeV. For values below the dotted line an initial value $\phi_{\rm PT} > f$ is required to reproduce the correct DM abundance.}
	\label{fig:inhomconstraints}
\end{figure}

We consider an inhomogeneous dark magnetic field generated with $1\%$ of the plasma energy density\footnote{We take the total energy in the dark sector to be 10\% of the energy density of the Standard Model bath.  Since our results only depend on $B/f$, the only change that arises by decreasing this 10\% number, and hence $B$, is to increase $f$.  The effect of this combination is to move around the dotted lines, the region that requires clockworking, in figure \ref{fig:inhomconstraints}.} by a first order phase transition at $T_{\rm PT} > 1$ GeV and subsequently diluting as $b\propto a^{-2}$. Such phase transitions typically have bubble sizes $\sim 1/(100 H_{\rm PT})$~\cite{Caprini:2015zlo,Hindmarsh:2020hop} and thus generate inhomogeneities of order $k\sim 100 H_{\rm PT}$. Requiring the ALP be frozen during the phase transition, $H_{\rm PT} > m_\phi$, gives an upper bound on the ALP mass for each choice of $T_{\rm PT}$. To evade cosmological bounds, and prevent suppressing the matter power spectrum at small scales, we also require the vector to dilute like matter sufficiently early. To be conservative, we require that the produced dark photons are non-relativistic before $T=10$ keV. Inspection of equation \eqref{eq:GlidingFormulaInh} shows that this gives two conditions 
    \begin{eqnarray}
        &&T_M > 10 \, \text{keV},\label{eq:Cond1}\\
        &&m_{\rm eff}(10 \, \text{keV}) \approx m_A.\label{eq:Cond2}
    \end{eqnarray}
The first condition gives a lower bound on $m_\phi$. For small values of $f$ the lower bound will meet the upper bound and close out the parameter space, so this condition effectively places a lower bound on $f$. The second condition, \eqref{eq:Cond2}, gives a lower bound on the dark photon mass. Ensuring the vector is the lighter field at the conversion time 
\begin{equation}
    m_{\rm eff}(T_M)<m_\phi, 
\end{equation}
places a lower bound on $m_\phi$ which rises with $f$. This effectively places an upper bound on $f$ since for large $f$ the upper and lower bound on $m_{\phi}$ will meet. Typically the lower bound on $f \sim 10^8$ GeV and the upper bound is $f\sim 10^{12}$ GeV. 

We vary $f$ and determine the allowed parameter space for the dark photon to constitute the entirety of dark matter. The allowed region of parameter space is shown in figure \ref{fig:inhomconstraints}. The diagonal line comes from requiring $m_A<m_\phi$. 
For a wide range of $\phi$ masses it's possible to produce dark photon dark matter down to $m_A \sim 10^{-13}$ eV. It's easier to produce higher masses by having the phase transition occur at higher temperatures, but pushing for masses any lower than this is hard. We can produce ligher dark photons by having the phase transition occur later since $k$ is related to the Hubble scale at the time of the phase transition, the lower the temperature the smaller $k$ will be. However, once we go much below $\sim 1$ GeV the upper and lower bound on $f$ quickly meet leaving no parameter space. 

Finally, as mentioned in section \ref{sec:Gliding}, for the conversion to be efficient the dark photon must be lighter when $b = m_\phi$,
\begin{equation}
    m_{\rm eff}(T_M)<m_{\phi}.
\end{equation}
This condition forces us to consider $\phi_{\rm PT} > f$ in order to achieve the right abundance for some regions of parameter space, which would likely require using the clockwork mechanism \cite{Kim:2004rp,Kaplan_2016}. The initial misalignment for an ALP is typically $\phi_{\rm PT} < f$, so the initial energy density in the ALP is $\rho_{\rm PT} \lesssim m_\phi^2f^2$. In order to reproduce the correct abundance for lower masses we can simply increase $f$. However, for fixed $B$, as $f$ increases the mixing becomes unimportant earlier, and thus $T_M$ increases. Since the effective mass decreases with time, this means that the effective mass at $T_M$ also increases. This gives a lower bound on $m_{\phi}$ which rises with $f$, and eventually closes out the parameter space. The dotted lines in figure \ref{fig:inhomconstraints} show the boundary between regions of parameter space that require clockwork to reproduce the observed dark matter abundance, and those that don't. For each phase transition temperature, the region below the dotted line requires $\phi_{\rm PT} > f$.


\section{Conclusion} \label{sec: concl}

In this article, we demonstrated how an axion like particle could be adiabatically converted into a dark photon in the presence of a magnetic field. This provides a new production mechanism for light vector dark matter, which can explain the dark matter abundance for dark photon masses as small as $10^{-13}$ eV. This mechanism differs from other scenarios in which an ALP energy density is transferred to dark photons in two relevant aspects. Firstly, the momentum of the newly produced dark photons is not directly related to the mass of the ALP, and is instead only sensitive to the scale of inhomogeneity of the magnetic field. Secondly, even though the transfer is not through a decay of the ALP, the leftover ALP abundance is exponentially smaller than the produced dark photon, due to the adiabaticity of the conversion process.

Throughout the paper, we considered only the simplest of scenarios and it would be interesting to see what occurs when these assumptions are relaxed.  One of the critical assumptions was that whatever sector the B field was a part of only had charged particles whose masses are sufficiently large to not change the evolution of the B field during the conversion time.  It would be interesting to see if the mechanism proposed in this paper was able to be extended to cases where this wasn't true, as would be the case if the B field belonged to the standard model photon. In addition, one of the main limitations to extending the mechanism to smaller dark photon masses was due to producing dark matter that is too warm. Thus, it would also be interesting to explore other production mechanisms of near homogeneous magnetic fields which would produce dark photons nearly at rest.  Next, a more realistic modeling of the magnetic fields resulting from a first order phase transition is important.  While we expect all of our results to follow through in this case, it is not guaranteed.  Finally, like all dark photon production mechanisms in the small $m_A$ limit, our approach requires living deep in the Stuckelberg limit.  The presence of the dark Higgs ($m_H \lesssim T_{\text{PT}}$) introduces a plethora of problems.  The dark Higgs would result in a large plasma mass suppressing conversion, string forming instabilities~\cite{East:2022ppo,East:2022rsi}, as well as often being more observable than the dark photon~\cite{Davidson:2000hf,Vinyoles:2015khy,Fung:2023euv}.  It would be interesting if there existed a production mechanism that circumvented these constraints.

\section*{Acknowledgments}
We thank Yuhsin Tsai for collaboration on the early stages of this project and useful comments on the draft.  
EB and AH are supported by NSF grant PHY-2210361 and the Maryland Center for Fundamental Physics. S.D. is partly supported by the McDonnell Center for the Space Sciences.

\appendix

\section{Constraints in a homogeneous magnetic field}
\label{app:homogeneous constraints}

\subsection{Conditions for Full Conversion}

In section \ref{sec:homogeneous} we used the approximation that all the energy was in the instantaneous slow mode, with the justification that the initial conditions excite mostly the slow mode. There are two ways we can end up with some of the energy in the fast mode
\begin{enumerate}
    \item Firstly, the initial conditions \eqref{eq:cosmoinitial1}, and \eqref{eq:cosmoinitial2} will mostly excite the slow mode, but there is some initial coefficient in the fast mode.
    \item Secondly, some initial slow mode may convert to fast mode due to the time dependence.
\end{enumerate}
There is a quick argument that we may neglect any initial fast mode amplitude. At the initial time, $t_{\text{PT}}$, both masses are small relative to both the mixing and the friction. We study the equations of motion \eqref{eq:EOM1}
in the massless limit. In this limit the slow mode has zero frequency. Making the substitutions $X=a^{5/2}\dot{\phi}$, $Y=a^{3/2}\dot{A}$ leads to equations of motion
\begin{equation}
    \begin{split}
        &\dot{X} + \frac{1}{2} H X = b Y,\\
        &\dot{Y} + \frac{1}{2} H Y = - b X.
    \end{split}
\end{equation}
This is a first order equation, and the solution is an oscillator with a constant amplitude 
\begin{equation}
    X, Y = C_{X,Y} e^{i\int dt' \omega(t')},
\end{equation}
where the frequency is $\omega_f = \sqrt{b^2 - \frac{1}{4}H^2}$. The energy density then falls off quickly as a function of the scale factor
\begin{equation}
\begin{split}
    \rho_f &\sim \dot{\phi}^2 + \dot{A}^2/a^2\\
    &= \frac{1}{a^5}(X^2+Y^2).
\end{split}
\end{equation}
Therefore any initial fast mode amplitude will quickly decay away and can be neglected.

\begin{figure}[ht]
\centering	\includegraphics[width=0.8\linewidth]{"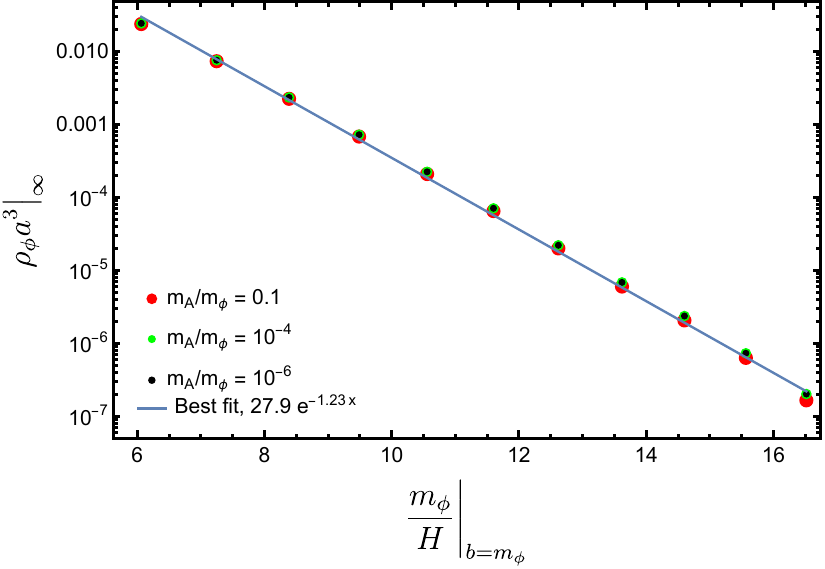"}
	\caption{We plot the energy density left in the scalar at late times against the quantity inside the exponential in equation \eqref{eq:LZfinal} for various choices of vector mass, and found the exponential scaling to be a good fit.}
 \label{fig:LZ}
\end{figure}

Assuming the fields are initially in the slow mode, due to the adiabatic evolution of the time-dependent quantities after the oscillations begin, it is natural to guess that the probability of transitioning to the fast mode would have the same behaviour as in the Landau-Zener problem \cite{Landau,Zener}
\begin{equation}
    \mathcal{P}(\textrm{slow} \rightarrow \textrm{fast}) \propto \exp\biggl(- c \frac{b^2}{\frac{d}{dt}(\omega_f-\omega_s)}\biggr),
\end{equation}
where the quantities in the exponential are to be evaluated at $t_M$. Up to order one coefficients, $\dot{\omega}_{iM} = \dot{b}_M$. Since at late times the fast mode is the scalar and the slow mode is the vector, the transition probability will scale the same way as the ratio of energy densities
\begin{equation}
\label{eq:LZfinal}
    \frac{\rho_\phi}{\rho_{\rm tot}} \propto \exp\biggl(-c \frac{b_M^2}{\dot{b}}\biggr) = \exp\biggl(-c \frac{m_\phi}{H_M}\biggr),
\end{equation}
where we used that the b-field dilutes with the scale factor, $\dot{b} \propto b H$, and absorbed all order one coefficients into the constant $c$. We found numerically that the guess based on Landau-Zener fits the data, and figure \ref{fig:LZ} shows some examples of testing the exponential scaling.

\subsection{Radiation Like Regime}
In this section we study the regime neglected in the main text with
\begin{equation}
    b(t_X)\leq \mphi.
\end{equation}
That is, the scenario when the mixing becomes unimportant before the slow mode begins oscillating $t_M < t_X$.

Somewhat surprisingly, in this regime, the energy density still goes into the slow mode as long as equation \eqref{eq:LZ} is still satisfied.  After mixing becomes unimportant, instead of the misaligned scalar depositing energy into the heavy oscillating scalar mode, it deposits its energy into the kinetic energy of the slow mode $A_D$.  Unlike a kinetic energy dominated scalar, which dilutes away as $a^{-6}$, a kinetic energy dominated vector dilutes away as $a^{-4}$.  Eventually, the mass term becomes important and the slow mode transitions to behaving as a non-relativistic vector.

\begin{figure}[t]
\centering
	\includegraphics[width=0.7\linewidth]{"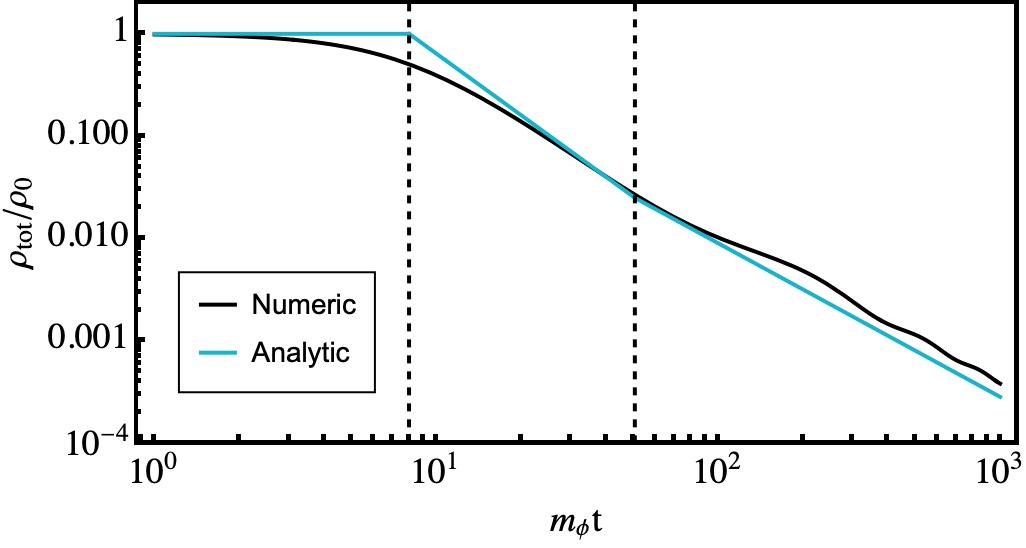"}
	\caption{The total energy density in the radiation like regime matches the approximation in \eqref{eq:rad}, that the energy density is frozen until the mixing becomes unimportant, at which point it dilutes like radiation until the crossing time, after which it dilutes like matter, with the all the energy in the dark photon. The vertical lines indicates the characteristics times, $t_M$ and $t_X$.}
	\label{fig:Homrad}
\end{figure}

In more detail, the same approximation as section \ref{sec:homogeneous}, $m_\phi^2\phi \approx b/a \dot{A}_D$ implies
\begin{equation}
    \frac{1}{2}m_\phi^2 \phi^2 \approx \frac{\dot{A}_D^2}{2a^2}.
\end{equation}
$\phi\approx \phi_{\rm PT}$ because the field is frozen until the time $t_M$, from which it's clear that the total energy at this time is $\rho_{\rm tot}(t_M) \approx m_\phi^2\phi_{\rm PT}^2$.  At time $t_M$, the scalar deposits all of its energy into the slow mode.  In the time $t_M < t < t_X$, equation \eqref{eq:GlidingEOM2} is no longer solved by the WKB approximation because friction is still larger than the slow frequency. Since the mixing has fallen off, $\lambda \ll 1$, equation \eqref{eq:GlidingEOM2} becomes
\begin{equation}
    \ddot{A}_D + H\dot{A}_D \approx 0\implies \dot{A}_D \propto \frac{1}{a}.
\end{equation}
Up until the crossing time the energy is therefore diluting like radiation
\begin{equation}
    \rho_{\rm tot} \approx \rho_A \approx \frac{1}{2a^2}\dot{A}_D^2 \propto \frac{1}{a^4},
\end{equation}
with the energy density in $\phi$ falling much faster than this.
After the crossing time the mass term becomes important and the vector dilutes like matter, so the energy density in dark matter at late times is given by
\begin{equation}
    \rho_A(t) \approx m_\phi^2\phi_{\rm PT}^2 \left(\frac{a_M}{a_X}\right)^4\left(\frac{a_X}{a(t)}\right)^3.
\end{equation}
In figure \ref{fig:Homrad} we plot the total energy density against the following approximation
\begin{equation}
    \rho_{\rm tot}(t) = \begin{cases}
        m_\phi^2\phi_{\rm PT}^2, & t<t_M\\
        m_\phi^2\phi_{\rm PT}^2 \left(\frac{a_M}{a(t)}\right)^4, & t_M < t < t_X\\
        m_\phi^2\phi_{\rm PT}^2 \left(\frac{a_M}{a_X}\right)^4\left(\frac{a_X}{a(t)}\right)^3, & t > t_X
    \end{cases}\label{eq:rad}
\end{equation}

\subsection{Constraints for the homogeneous case}
We show the constraints on the mass range of the dark photon if we assume that the magnetic field can be treated as homogeneous, and the dark photon constitutes the entirety of dark matter. We assume that a magnetic field was generated during a first order electroweak phase transition, with a percent of the universe total energy density. The phase transition occurs at a temperature $T_{\rm PT} \sim 100$ GeV. During this period the universe was radiation dominated. To a good approximation we assume that at temperature $T$ the energy density of the universe is given by the relativistic particles $m \ll T$,
\begin{equation}
    \rho = \frac{\pi^2}{30}g_*(T) T^4,
\end{equation}
where $g_*(T)$ counts the number of effectively relativistic degrees of freedom in the standard model. As in the main text we take the conservative $T_{\rm DM} = 10$ keV. From now on anything with a subscript DM is evaluated at this time/temperature. By $T_{\rm DM}$ we must have generated the dark photons, $T_X>T_{\rm DM}$, and they must be acting like matter, $b_{\rm DM}<m_\phi.$

\begin{figure}[ht]
\centering
	\includegraphics[width=0.6\linewidth]{"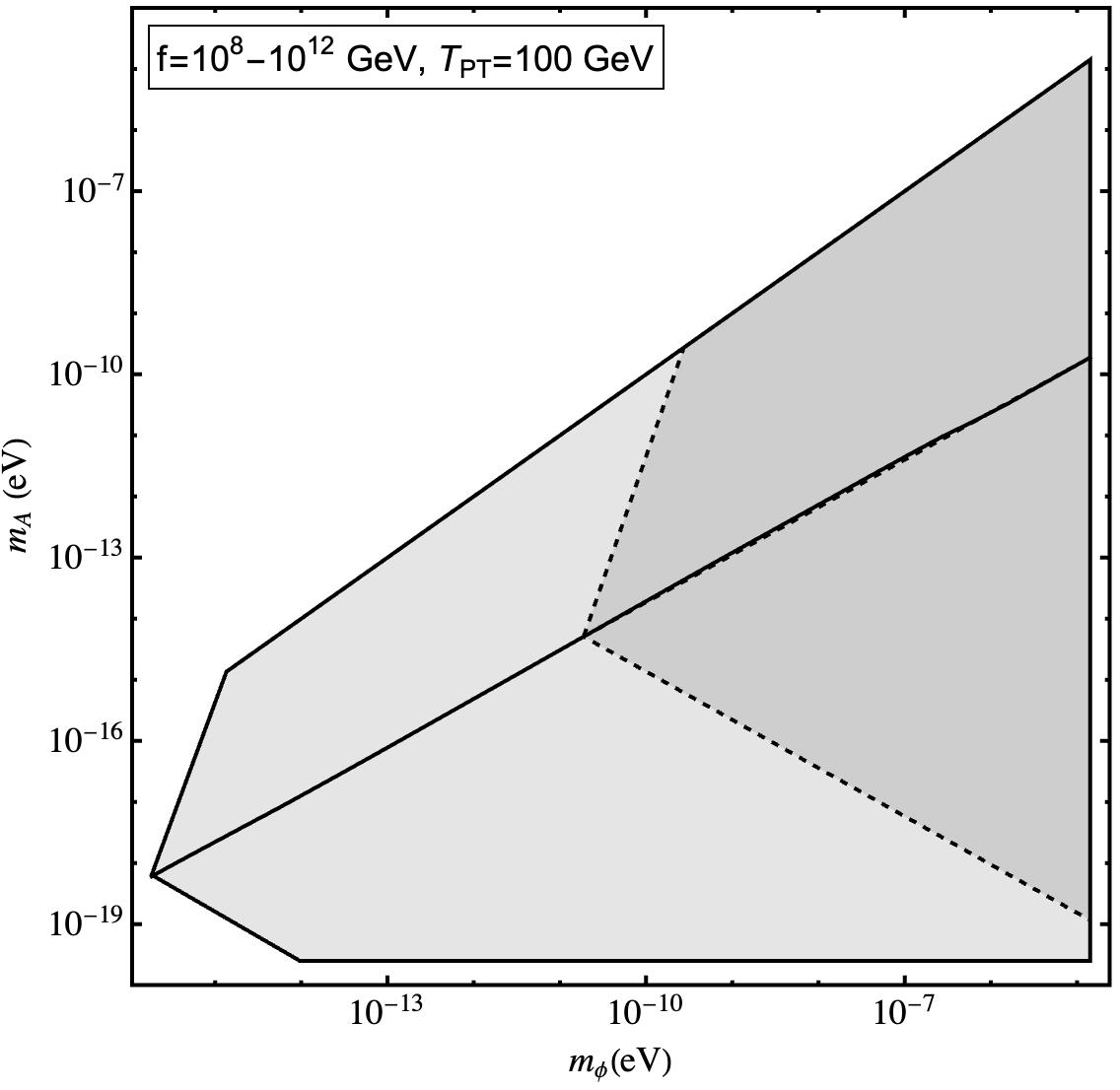"}
	\caption{The constraints on the particle masses, with the allowed region shaded in grey. The upper box is the gliding regime whilst the lower box is the radiation like regime. The area to the right of the dotted line can produce the correct dark matter abundance for $\phi_{\rm PT} < f$.}
	\label{fig:homconstraints}
\end{figure}

Supernova constraints give $f\gtrsim 10^{8.5}$ GeV \cite{Hook:2021ous} for a SM B field. With the given range of decay constants we want to know what values we can pick for $m_\phi$, and $m_A$. In figure \ref{fig:homconstraints} we plot the region of masses allowed with the previously mentioned constraints. The boundary conditions require $H_0>m_\phi$ which gives the vertical line on the right. With the more conservative choice of $T_{\rm DM} = 10$ KeV a homogeneous magnetic field can generate dark photons with a lower bound of $m_A\gtrsim 10^{-20}$ eV. We also note that the mechanism allows for a large separation of scales between the masses of the two particles, with the far right giving over 10 orders of magnitude separation. 



\section{More general inhomogeneities}

In the main text we studied the following field profile
\begin{equation}
    \label{eq:InhomogeneousField2}
    \vec{b}(x,t) = b_{\rm PT}\left(\frac{a_{\rm PT}}{a(t)}\right)^2\cos (kx) \hat{z},
\end{equation}
and showed that the inhomogeneity of a single mode can be modeled as an effective mass for the dark photon as in eq \eqref{eq:meff}. We would like to say this approximation holds even if the b-field is a wave-packet with its Fourier transform sharply peaked around the characteristic wave number, $k$.

To simulate in a finite spatial box with periodic boundary conditions we consider the following profile
\begin{equation}
\begin{split}
    \label{eq:InhomogeneousField3}
    \vec{b}(x,t) &= b_{\rm PT}\left(\frac{a_{\rm PT}}{a(t)}\right)^2\frac{1}{\sqrt{N^2+2+2N^{-2}}}\left[\frac{1}{N}\cos (kx/3+\phi_1)+\cos(kx/2 + \varphi_2)+N\cos(kx)\right.\\
    &\left.+\cos(2kx + \varphi_3)+\frac{1}{N}\cos(3kx + \varphi_4)\right]\hat{z},
\end{split}
\end{equation}
where $\varphi_i$ are randomly generated phases. If $N=1$ then the wave-packet receives equal contributions from all 5 modes. However, as we increase $N$ the wave-packet becomes more sharply peaked around the characteristic wave number, $k$.

\begin{figure}[t]
  \centering
   \subfloat{\includegraphics[width=0.32\textwidth]{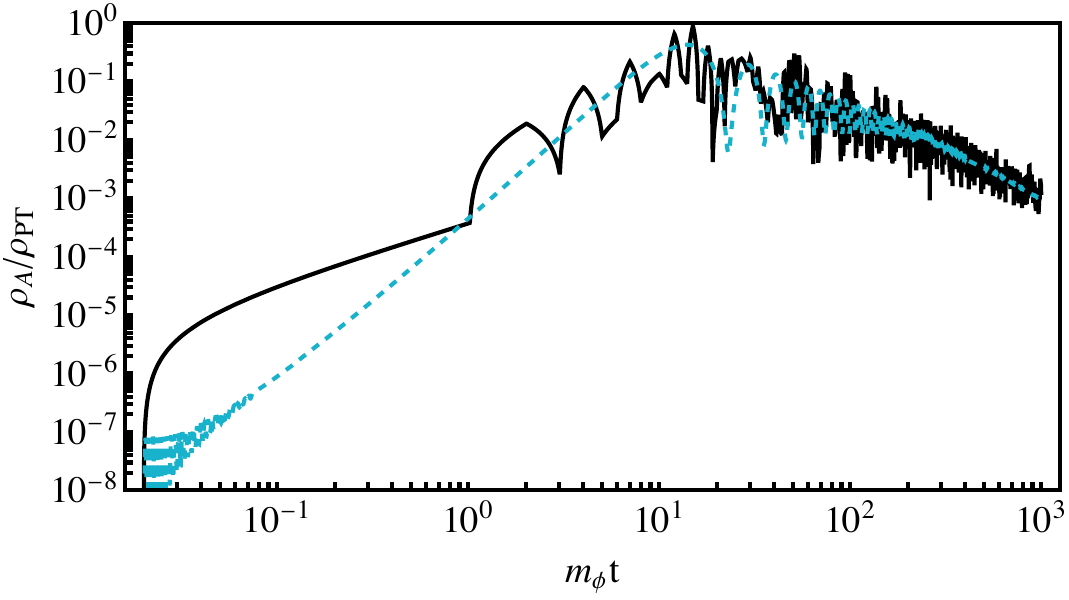}
  }
  \hfill
  \subfloat{\includegraphics[width=0.32\textwidth]{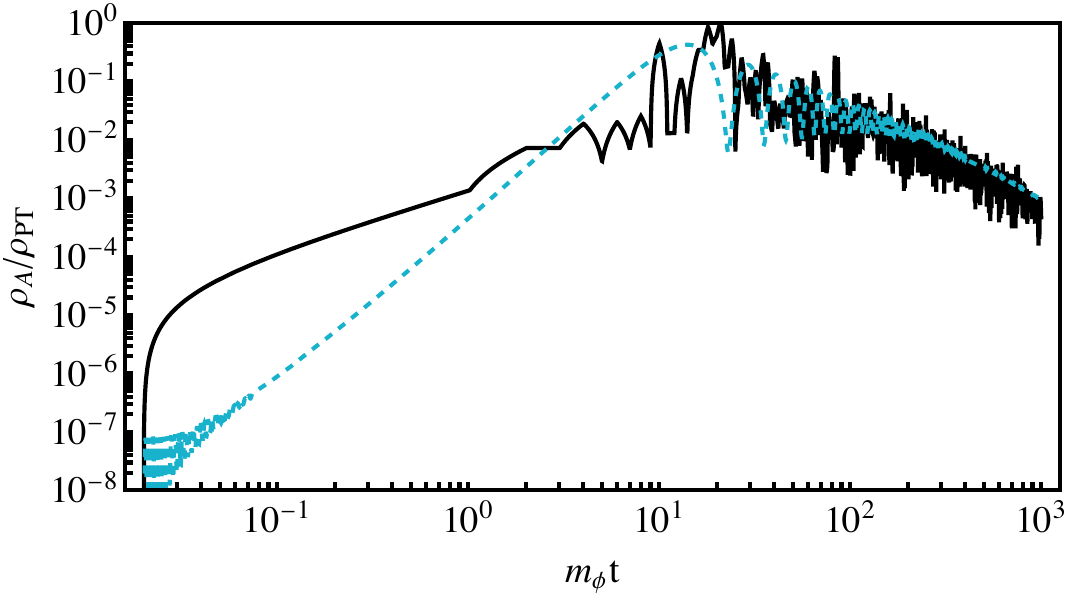}}
  \hfill
  \subfloat{\includegraphics[width=0.32\textwidth]{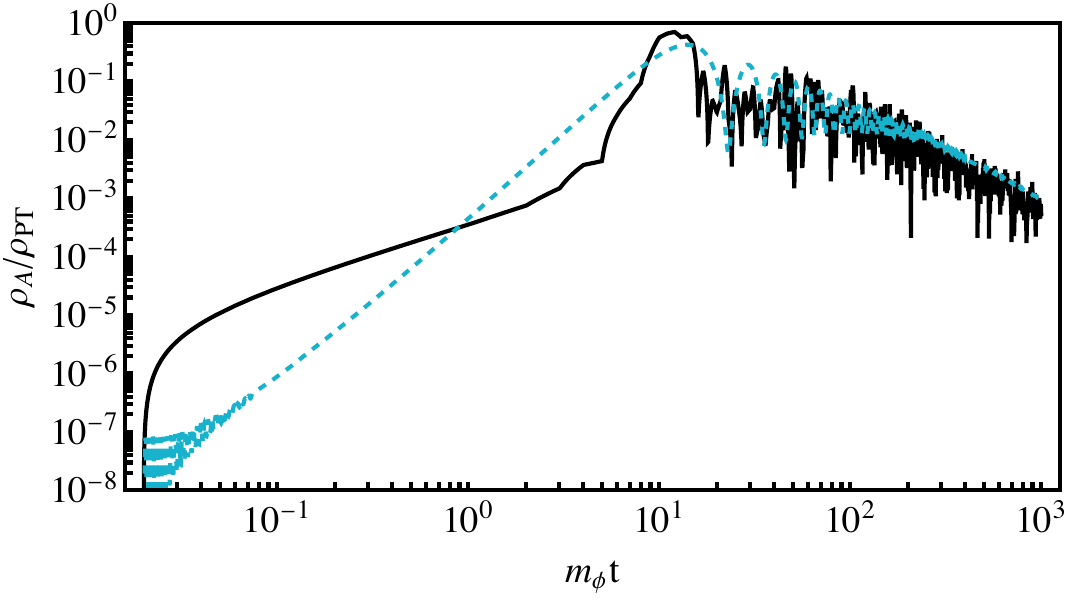}}
  \caption{A comparison of the vector energy density for the inhomogeneous field and the effective mass approximation, with $N=1,2,3$ from left to right. The black lines are the numerical approximation, and the blue line is from the effective mass approximation. In all simulations $m_\phi/m_A = 10$}
  \label{fig:EffectiveMassDemonstrationN}
\end{figure}

Figure \ref{fig:EffectiveMassDemonstrationN} shows the results for various values of $N$. We found that even for $N=1$ the effective mass approximation is still working. We expect that the conversion mechanism still works for more general magnetic field profiles, although more detailed simulations are required to verify this claim.

\bibliographystyle{JHEP}
\bibliography{references.bib}

\end{document}